\documentclass[12pt,preprint]{aastex}

\shorttitle{Mid-IR observations of LBGs}
\shortauthors{Barmby et al.}

\begin{document}

\title{Deep mid-infrared observations of Lyman-break galaxies}

\author{P. Barmby\altaffilmark{1}, 
J.-S. Huang\altaffilmark{1}, 
G.G. Fazio\altaffilmark{1}, 
J.A. Surace\altaffilmark{2}, 
R.G. Arendt\altaffilmark{3},
J.L. Hora\altaffilmark{1}, 
M.A. Pahre\altaffilmark{1}, 
K.L. Adelberger\altaffilmark{4},
P. Eisenhardt\altaffilmark{5}, 
D.K. Erb\altaffilmark{6}, 
M. Pettini\altaffilmark{7},
W.T. Reach\altaffilmark{2},
N.A. Reddy\altaffilmark{6}, 
A.E. Shapley\altaffilmark{8},
C.C. Steidel\altaffilmark{6}, 
D. Stern\altaffilmark{5},
Z. Wang\altaffilmark{1}, 
S.P. Willner\altaffilmark{1}}

\altaffiltext{1}{Harvard-Smithsonian Center for Astrophysics, 60 Garden Street, Cambridge, MA 02138; 
pbarmby@cfa.harvard.edu}

\altaffiltext{2}{Spitzer Science Center, MS220-6, California Institute of Technology, 
Pasadena, CA 91125}

\altaffiltext{3}{Science Systems and Applications, Inc.;
Code 685, NASA Goddard Space Flight Center, Greenbelt, MD 20771}

\altaffiltext{4}{Observatories of the Carnegie Institution of Washington, 813 Santa Barbara St.,
Pasadena CA, 91101}

\altaffiltext{5}{Jet Propulsion Laboratory, California Institute of Technology, 
Mail Stop 169-327, Pasadena, CA 91109}

\altaffiltext{6}{California Institute of Technology, MS105-24, Pasadena, CA 91125}

\altaffiltext{7}{Institute of Astronomy, Madingley Road, Cambridge CB3 0HA, UK}

\altaffiltext{8}{Astronomy Department, 601 Campbell Hall, University of California, Berkeley, CA 84720}

\begin{abstract}
As part of the In-Orbit Checkout activities for the {\it Spitzer Space Telescope}, 
the IRAC team carried out a deep observation (average integration time $\sim8$ hours)
of a field surrounding the bright QSO HS 1700+6416.  This field contains 
several hundred $z\sim 3$ Lyman-break galaxy 
candidates, and we report here on their mid-infrared properties, including
the IRAC detection rate, flux densities and colors, and the results of fitting
population synthesis models to the optical, near-infrared, and IRAC 
magnitudes. The results of the model-fitting show that previous optical/near-infrared 
studies of LBGs were not missing large, hidden old stellar populations.
The LBG candidates' properties are consistent with those of massive, star-forming 
galaxies at $z\sim3$. Other IRAC sources in the same field have similar properties,
so IRAC selection may prove a promising method of finding additional high-redshift galaxies.
\end{abstract}


\section{Introduction}
Observations of galaxies at high redshift provide important tests of
theories of galaxy and structure formation: galaxy masses, star
formation rates, and the distribution of these quantities are all
vital information.  Some of the best-known high-redshift galaxies are
the $z\sim3$ galaxies discovered with the Lyman-break technique
\citep{sh93}. These galaxies are forming stars rapidly and
are generally thought to be massive enough to be the
ancestors of today's large galaxies \citep{kla98,ccs98, ww03}. Rest-frame
near-infrared observations primarily sample the old stellar population
in galaxies and are not severely affected by dust extinction. They
should provide a less biased estimator of the galaxy mass than optical
measurements, which are sensitive to both internal extinction and
emission from young stars which make up little of the galaxy
mass \citep{be00,kc98}.  The four bands of the Infrared Array Camera
\citep[IRAC; ][]{irac} on the {\it Spitzer Space Telescope} \citep{sirtf}
probe the rest-frame near-IR at redshifts $z=2-4$.

This paper reports {\it Spitzer}/IRAC 3--10 $\mu$m observations of a 
Lyman-break galaxy field. The goals of the observations 
were to confirm that IRAC could detect $z\sim3$ galaxies and to characterize
the noise performance in the approach to the confusion limit.
Galaxy number counts in this field are 
addressed in a separate paper \citep{irac_numc}. IRAC properties of 
$z\sim2$ galaxies in this field are also addressed in a separate paper 
(Steidel et al.\ 2004, in preparation).
Throughout this paper we assume a cosmology with $H_0=70$~km~s$^{-1}$~Mpc$^{-1}$, 
${\Omega}_m =0.3$, and ${\Omega}_{\Lambda}=0.7$.

\section{Observations and data reduction}

The field observed was a Lyman-break galaxy field surrounding  
the $z=2.7$ QSO HS~1700+6416,
from the survey of \citet{lbg}. This particular field was chosen
because of its existing very deep optical photometry, 
which allowed selection of high-redshift galaxies, and its
position within the {\it Spitzer} Continuous Viewing Zone. The field also
includes two galaxy clusters (Abell~2246  at $z=0.25$ and an anonymous
cluster at $z=0.44$), whose X--ray emission
was discovered during a study of the QSO \citep{rei97}.
The IRAC observations were centered at coordinates
$17^{\mathrm h}01^{\mathrm m}17^{\mathrm s}$, 
$64^{\mathrm d}09^{\mathrm m}02^{\mathrm s}$ (J2000), about 3.7\arcmin\ southeast of the QSO.
The central IRAC pointing was offset in order to slightly lower the observed 
galaxy density and reduce the effects of confusion.
The two IRAC fields of view see different regions of the sky, so
the final 3.6 and 5.8~$\mu$m images extend further to the
southeast while the 4.5 and 8.0~$\mu$m images extend to the northwest.
The IRAC images were taken on (UT) 
14--16 October 2003 as part of {\it Spitzer} Program ID 620. 
We used 200-second frametimes in the 3.6--5.8~$\mu$m channels;
this observing mode uses 50-second
frametimes in the 8.0~$\mu$m band because of the higher sky background.
We obtained some additional data with 100-second frametimes,
but only used the 8.0~$\mu$m-band data (which again uses 50~s frametimes)
to avoid systematics in combining data taken with different frametimes.

The individual images were processed with the Spitzer
Science Center (SSC) pipeline,
then had their astrometric positions refined with 2MASS sources.
The zero levels of the 5.8$\mu$m-band images were adjusted to 
minimize differences in the overlap areas using the
overlap consistency option in the SSC processing tools.
The images were combined into mosaics with our IDL-based code
which re-projected the images to correct for distortion and combined them
with a sigma-clipping algorithm. The resulting mosaics are in the original
array orientation, with a pixel scale half as large as the 
original.\footnote{The mosaics will be available via
\url{http://cfa-www.harvard.edu/irac/publications}.}

\begin{figure}
\caption{IRAC images of the Q1700+64 field in the 4 bands (wavelength
labels in the upper right of the images). These images show
a portion (3.5\arcmin $\times$ 3.5\arcmin) of the area covered by all 
four IRAC bands; north is to the upper right and east to the upper left.
White circles indicate LBGs detected in all four IRAC bands; black circles 
are LBGs detected in three bands; white squares are LBGs detected in two bands.
\label{fig-mosaics}}
\plotone{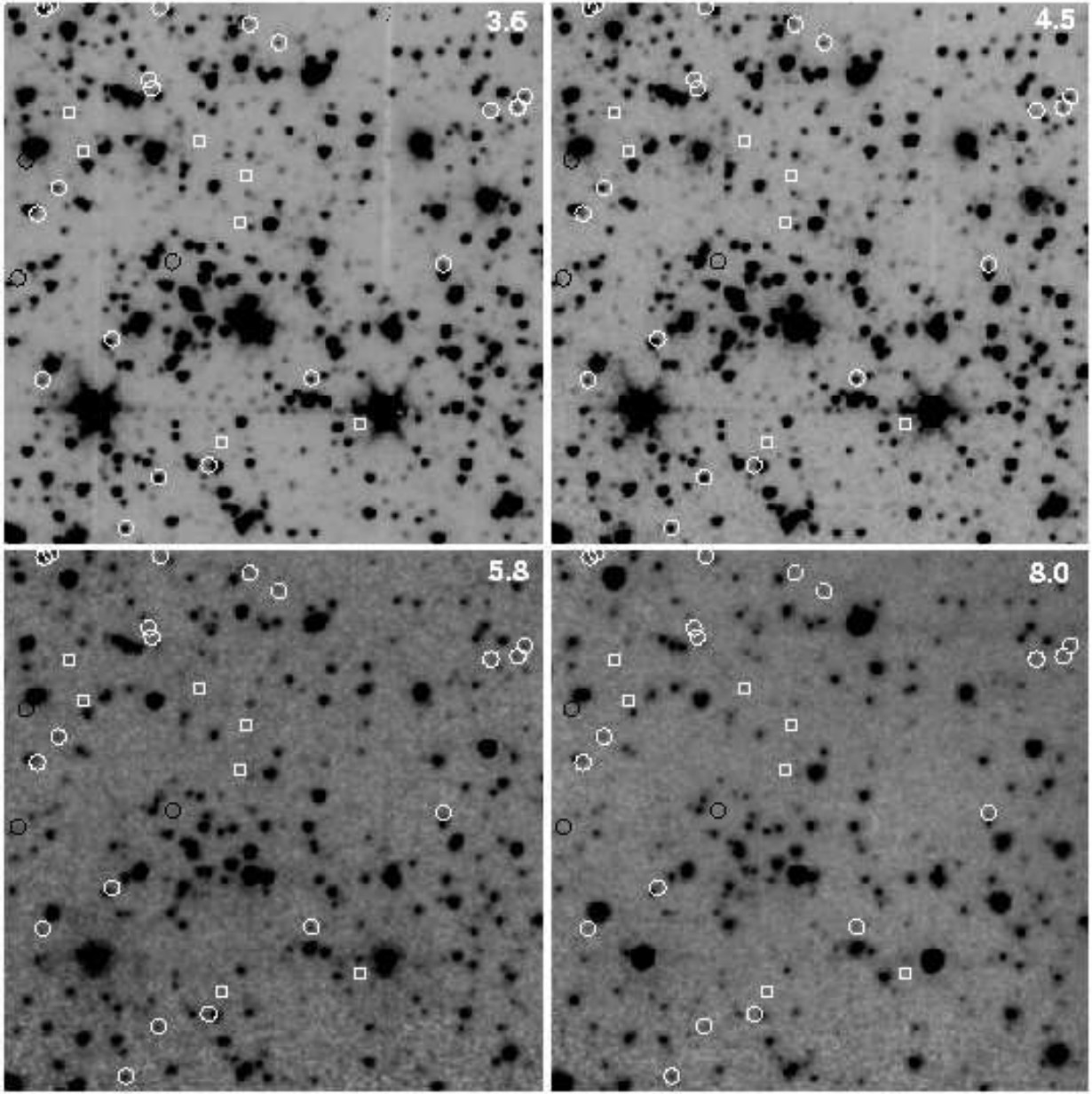}
\end{figure}

Figure~\ref{fig-mosaics} shows portions of the central regions of
the mosaic images in the 4 IRAC bands. Because of dithering, the effective exposure 
time across the mosaics varies, ranging from about 11 hours in the deepest
areas to 20 minutes near the edges. The average exposure time over 
the central $5\times10$ arcmin area is 7.8 hours.
The detection limits vary with the effective exposure time; approximate 
$5\sigma$ limits in the central regions of the images are 
0.45, 0.45, 0.8, and 0.9~$\mu$Jy (AB magnitudes of 25.0, 25.0, 24.2, 24.1)
in the four bands.

The optical data used for LBG selection are described by \citet{lbg}. Although
these data were obtained for a survey of $z = 2$ galaxies, the imaging
methods are the same for both $z=2$ and $z = 3$ selection (only the color
criteria differ). Briefly, the optical imaging observations were obtained at the William Herschel
4.2m telescope in 2001 May and supplemented with $U$-band images from
Keck I/LRIS-B later that year for the central region near the QSO.
The $1\sigma$ surface brightness limits in the 3 bands (in a 1\arcsec\ aperture)
are $U_n(AB)=28.9$, $G(AB)=29.0$, and $R_s(AB)=28.2$. 
A deep $K_s$-band image with $K_{AB}=23.9 (5\sigma)$ covering the central $9\times9$\arcmin\ 
was obtained at the Palomar Hale Telescope in 2003 June. Optical spectroscopy
of an LRIS-B mask targeting 22 $z\sim 3$ candidates was obtained
in 2003 September. Reduction followed the method described by \citet{ccs03}, and
17 redshifts were successfully measured, with $\bar{z}=2.911$.
The spectroscopically-confirmed galaxies had a median $R_s(AB)=24.6$.

\section{Analysis}

\subsection{Object detection and photometry}

Lyman-break galaxy selection was done using the $U_nG\cal R$ technique
described by \citet{ccs03}. A total of 445 LBG candidates in the
$15\arcmin \times15$\arcmin\ area centered on the QSO were selected,
to a limiting total magnitude of $R_s(AB)=25.5$.   These
objects are typically $>10\sigma$ detections in $\cal{R}$,
$15-20\sigma$ in $G$, and very faint or undetected in $U_n$.
In general, spectroscopic observations of LBG candidates show that
about 95\% are likely to be {\em bona fide} 
$z\approx3$ galaxies \citep{ccs03}. There are about 110 LBG candidates in the 
3.6 $\mu$m/5.8$\mu$m field of view, about 175 candidates in the 4.5$\mu$m/8.0$\mu$m 
field of view, and about 90 in the overlap area.
Parts of the final 3.6 and 5.8 $\mu$m images
extend outside the $U_nG\cal{R}$ detection images.

To detect LBG candidates on 
the IRAC images, we used the image world coordinate system
to project the candidates' positions
into the image plane and performed aperture photometry at those positions.
We experimented with allowing the photometry algorithm to re-center the objects;
if the objects moved more than 1\farcs2 in either $x$ or $y$ (indicating that the
identification was in danger of moving to a nearby object), we used the
un-centered position. This is important because the IRAC images are crowded, particularly 
at the shorter wavelengths (the IRAC point spread function FWHM
is about 1\farcs8--2\farcs0).
As an indicator of detection significance, we used the
IRAF/APPHOT photometric uncertainty inside a 1\farcs5 radius aperture, with sky 
counts determined in 15--20\arcsec\ annuli. Objects with magnitude 
uncertainty~$<0.217$~mag ($S/N>5$) were considered detected (see 
Figure~\ref{fig-mosaics} for examples).
The final IRAC detection numbers
for the 4 bands are given in Table~\ref{detnum}. Most of the LBG
candidates were detected at 3.6 and 4.5~$\mu$m; the 
detection rate decreases with increasing wavelength to 45\%   
at 8.0$\mu$m. Photometry of the LBG candidates
was completed by applying aperture corrections from the 1\farcs5 measurement
aperture to the 12\farcs2 aperture used for flux calibration. The
corrections, derived from the measured IRAC PSF (S.T. Megeath, priv. comm.), 
are $-0.52$, $-0.55$, $-0.74$, and $-0.85$ mag.

To generate a sample of IRAC-selected non-LBG objects for comparison
to the LBG candidates, we used DAOFIND
to find objects to a threshold of $5\sigma$ above local background.
The detection images were the IRAC mosaics multiplied by the square
root of the coverage maps (to give uniform noise across the images)
with areas near the image edges, near bright stars, or near instrumental
artifacts masked out. Detection was
performed in the 3.6 and 4.5~$\mu$m band images only, with object positions
transferred to the frame of the 5.8 and 8.0~$\mu$m band images for photometry.
For  a more reliable sample, we included only objects detected in
both 3.6 and 4.5~$\mu$m bands, which restricts the area used to about
42 arcmin$^2$. About half of the LBG candidates in the field
were detected as part of this sample (the rest were 
too close to brighter objects or masked areas), and were removed.
The final non-LBG sample contained 1931 objects.
Photometry of the non-LBG sample
was done in an identical manner to that of the LBG candidates.
No attempt was made to separate stars and galaxies 
in the non-LBG sample on the basis of spatial extent; 
the size of the IRAC PSF makes this impractical for distant galaxies.
The galactic coordinates of the field ($l=94.3, b=+36.2$) are
such that we do not expect stellar contamination to be a serious
problem for faint objects.

\subsection{LBG properties: flux and color distributions}

To compare the optical properties of IRAC-detected and non-detected LBGs 
we used the optical catalog generated for LBG selection. 
There was no significant difference (K-S test probability $>$ 5\%) 
in $\cal R$ between objects detected and not detected in the 3.6~$\mu$m band.
However, there are so few non-detections in this band (only 10) that
this comparison is not particularly meaningful.
A K-S test of the $\cal R$ magnitude distributions of 
IRAC detections and non-detections
in the 4.5, 5.8, and 8.0~$\mu$m bands showed that the differences
in mean magnitudes are statistically significant.
The IRAC-detected
objects are roughly 0.2--0.4 mag brighter in $\cal R$ than
the non-detected objects.  
There is, however, no clear correlation between $\cal R$ and IRAC 
magnitudes for the IRAC-detected objects. 

\begin{figure}
\caption{Two-color plots of LBG candidates (asterisks), confirmed
$z\sim 3$ LBGs (filled squares) and non-LBG sample (dots). 
Lines are template SEDs for $0.5<z<4$, with redshifts
0.5,1,2, and 4 marked. The different templates are: Sbc spiral (thick solid), 
irregular (dashed), M82 (dotted). Thin solid lines indicate color-selection
criteria for high-redshift galaxies.
\label{twocolor}}
\plotone{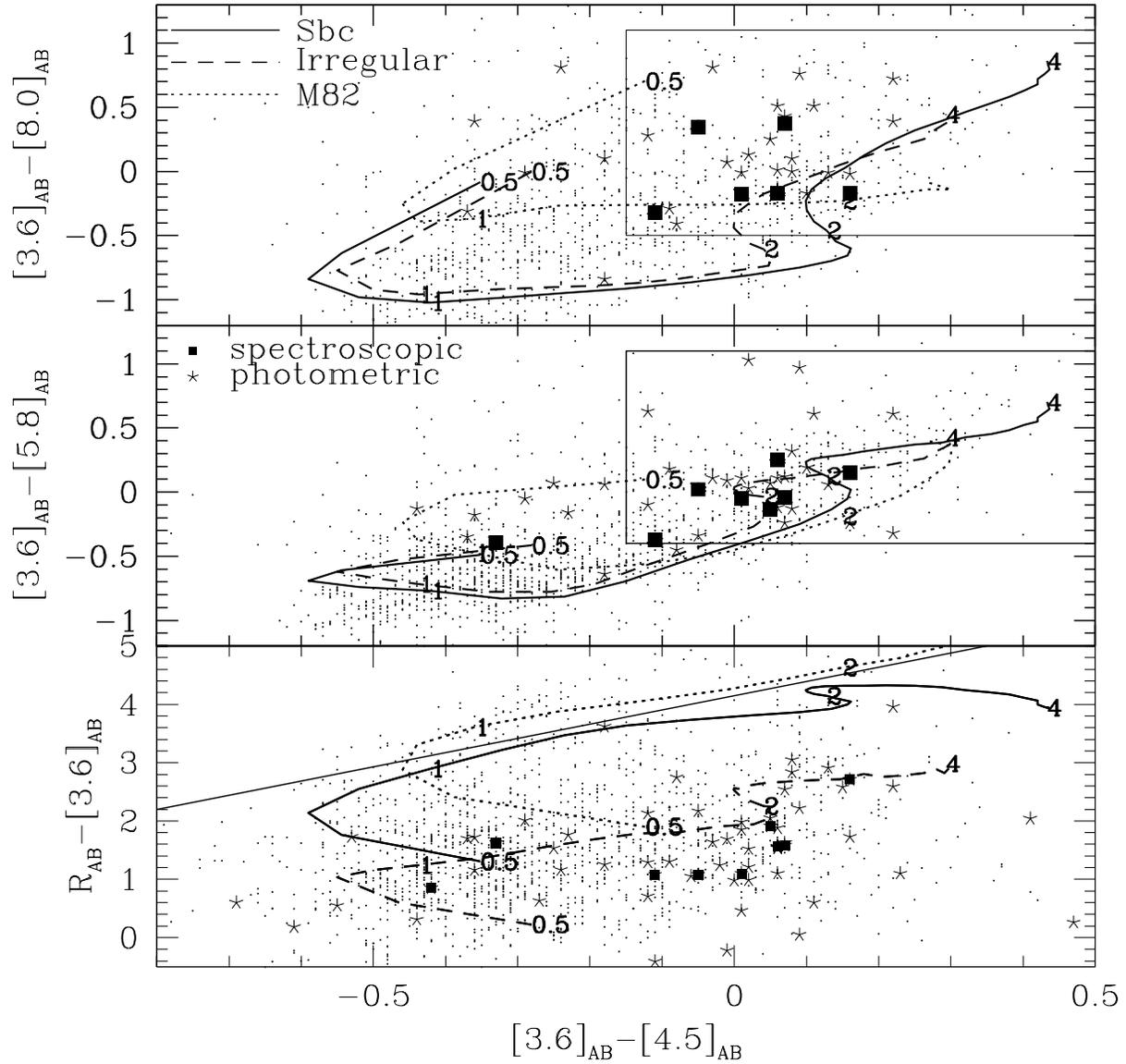}
\end{figure}

We can also compare the IRAC flux density distributions of the 
LBG candidates and the non-LBG sample. The LBGs have median IRAC 
magnitudes 0.5--0.7 fainter than the non-LBG sample. 
The spectroscopically-confirmed galaxies
are slightly brighter than the full candidate list 
in the 3.6 and 4.5 $\mu$m bands and  slightly fainter
in the longer-wavelength bands (but their numbers are small: only
six such galaxies have four IRAC-band detections).
The 33 LBG candidates detected in all four IRAC bands have
median AB magnitudes in the IRAC bands of 22.7, 22.7, 22.6 and 22.5
(flux densities in $\mu$Jy: 3.1, 3.1, 3.3, 3.8). 
Two-color plots are shown in Figure~\ref{twocolor}. 
The LBGs are redder than the comparison objects in [3.6]--[4.5], 
for which the color limit is constant with magnitude. Selection
effects bias the observed color distributions in [3.6]--[5.8] and
[3.6]--[8.0]: because of the different sensitivies in the
IRAC bands, bluer objects are observable only at bright [3.6] magnitudes.

Figure~\ref{twocolor} also shows predicted colors for template SEDs.
The templates combine optical and near-infrared data from the {\sc hyperz} 
package \citep{bmp00} and mid-infrared data from \citet{lu03}. 
As \citet{pdf01} report for optical/near-infrared data, 
the `irregular galaxy' template SED is the closest
match to the LBG candidates' colors, particularly the spectroscopically-confirmed
objects. 
Besides being known to be at the correct redshift, the spectroscopic galaxies
also tend to be brighter, so their colors should be better determined.
A substantial number of LBG candidates fall well away from the 
template colors, which could happen for several reasons: photometric error
(for example, due to contamination by nearby objects), incomplete or inapplicable templates.

Some objects in the comparison sample have the same fluxes and colors
as the LBG candidates, and therefore could be high-redshift galaxies.
Using the top two panels of Figure~\ref{twocolor}, we suggest selection
criteria for $z>2$ galaxies 
based on both the LBG candidate observations and the template tracks.
The criteria $[3.6]_{\rm AB}-[4.5]_{\rm AB}>-0.15$, 
$1.1>[3.6]_{\rm AB}-[5.8]_{\rm AB}>-0.4$, $1.1>[3.6]_{\rm AB}-[8.0]_{\rm AB}>-0.5$, 
and $[3.6]_{\rm AB}>22$ give good separation between 
low and high-redshift galaxy templates, while including about
two-thirds of the LBG candidates and all but one spectroscopically-confirmed LBG. 
This color criterion would also pick up some M82-like SEDs at $1<z<2$; however,
the bottom panel of the Figure shows that adding the criterion 
$R_{\rm AB}-[3.6]<4.1+2.4([3.6]_{\rm AB}-[4.5]_{\rm AB})$
should reduce the number of such objects.
These criteria define a subsample of 86 objects from the non-LBG sample,
giving an areal density of approximately 2~arcmin$^{-2}$. 
This is comparable to the number density of LBGs in the same
area (2.0~arcmin$^{-2}$). While the LBG technique is designed
to select galaxies over a fairly narrow redshift range, the use of
IRAC/optical colors may permit selection of high-$z$ galaxies 
without biases toward UV-bright objects, albeit with an as yet
undetermined contamination rate \citep[see also][]{lh}.
Confirmation of such color selection with spectroscopic redshifts is of course
necessary, and this will be the subject of future investigations.

\subsection{Spectral energy distributions and model fitting}

At high redshifts, the IRAC bands sample the rest-frame near-IR
light of galaxies. This light is expected to be a good tracer of
galaxy stellar mass since it comes mainly from low-mass older stars.
To test this assumption, we compared stellar population synthesis
to the combination of optical, near-infrared, and IRAC data.
Following~\cite{aes01}, we used the population synthesis models
of \citet{bc03} to generate synthetic spectral energy distributions
for composite stellar populations with solar metallicity, a 
\citet{sal55} initial mass function, and star formation rates (SFRs)
\begin{equation}
\Psi(t)={\Psi}_0 e^{-t/\tau_{\rm SFR}} \label{sfr}
\end{equation}
with $\tau_{\rm SFR} = 10, 50, 100$~Myr, and galaxy ages at $z=3$ of
10, 50, 100, 200, 500, 1000, 1200, 1500 and 2000~Myr,
corresponding to formation redshifts $3.21<z_f<30$.
As a matter of convenience, we used the \citet{bc03} code's internal
prescription for extinction, derived from \citet{cf00}. Models
with four different values of the extinction parameter ${\tau}_V$
(0.5, 0.75, 1.0, 1.5) were generated. Predictions for observed fluxes were made
from the model SEDs using the program {\sc cm\_evolution} included in
Bruzual \& Charlot's software distribution. 
We expect the rest-frame UV photometry to provide a more sensitive
estimate of the instantaneous SFR in these models, while the rest-frame
near-IR photometry from IRAC will provide the constraints on the
integrated stellar mass.

\begin{figure}
\caption{
Spectral energy distributions for spectroscopically-confirmed
LBGs and a composite of the photometric candidates. Low-resolution
SEDs for the best-fitting \citet{bc03} models are shown; models
are normalized to the data at $\lambda_{\rm obs}$= 3.6$\mu$m ($\lambda_{\rm rest}\approx 0.9 \mu$m).
From shortest to longest wavelength, the photometric points marked are in the 
$G$, $\cal{R}$, $K_s$, 3.6, 4.5, 5.8, and 8.0~$\mu$m bands. Ages and $\tau_{\rm SFR}$
are in Myr; stellar masses are in $M_\sun$.
\label{seds}}
\plotone{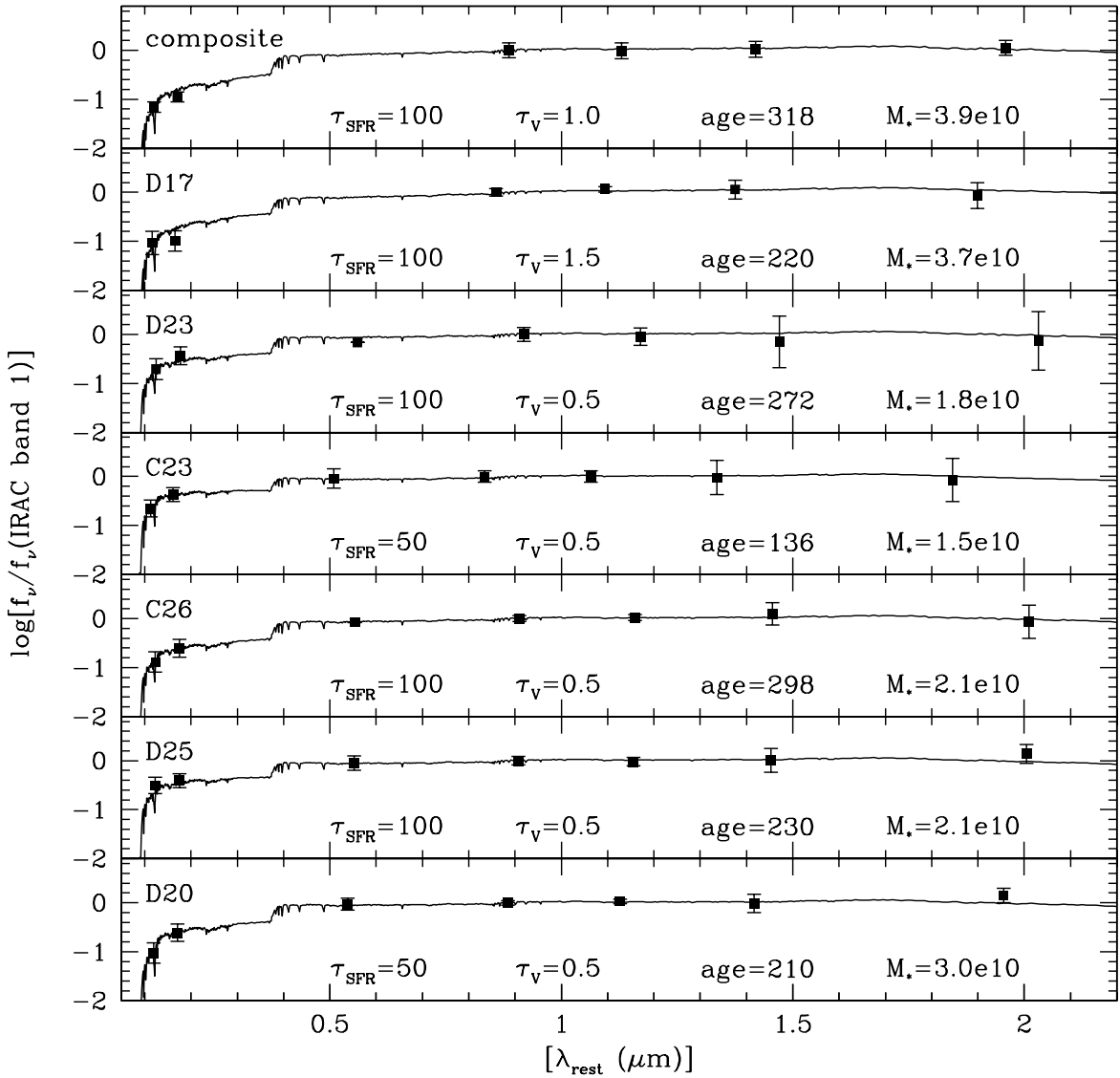}
\end{figure}

The best-fitting models were determined by minimizing ${\chi}^2$, 
the sum of the difference between model and observations. The 
\citet{bc03} models have a tabulated stellar mass for each age, so the 
overall best-fit normalization of the model apparent magnitudes
fixes the model stellar mass.
We fit models to the $GRK_s+$IRAC magnitudes of each of the six
spectroscopically-confirmed galaxies individually and to a `composite' 
SED made from the median fluxes of the the 26 unconfirmed candidates 
(assumed to be at $z=3.00$). The model SEDs and observations, with
model parameters, are shown in Figure~\ref{seds}. Current star formation rates
$\Psi(t_{\rm obs})$ are derived using Equation~\ref{sfr} and
requiring $\int_0^{t_{\rm obs}} \Psi(t) dt = (1-R)M_*$
\citep[$R=0.28$ is the fraction of formed stellar mass returned
to the ISM;][]{cole00}

Most of the observations are best fit by models with 
star formation $\tau_{\rm SFR} = 50-100$~Myr, relatively young ages
(100-300~Myr), and stellar masses in the range $1.5-4\times 10^{10} M_{\sun}$. 
The resulting star formation rates are $7-33 M_{\sun}$~yr$^{-1}$.
The largest SFR is for D17, a galaxy without a measured $K_s$-band
magnitude which also has a large best-fit extinction. This object
has bluer IRAC colors than the predicted SED of the prototype dusty starburst 
M82 but could be intermediate in type between M82 and a nearby irregular.
Previous results on SED fitting of LBGs with optical and near-infrared
observations are presented by \citet{aes01} and \citet{pdf01};
the latter used a much more extensive grid of population synthesis models.
Our model parameters are well within the range of
best-fit parameters found by these two groups, indicating that 
IRAC-detected LBGs are not particularly unusual members
of the LBG population, and the previous optical/near-infrared studies
were not missing large, hidden old stellar populations.
More sophisticated model-fitting, using a wider range of stellar population
parameters and extinction prescriptions, is clearly possible
but beyond the scope of this paper. Such analysis will be pursued
in subsequent work, using these data and a larger sample of LBGs 
in the Extended Groth Strip survey area.

\section{Summary}

Deep observations with IRAC detect nearly all $R_{\rm AB}< 25.5$
Lyman-break galaxy candidates at 3.6 and 4.5~$\mu$m, and about
half of the candidates at 5.8 and 8.0~$\mu$m. The color
and magnitude distributions of LBGs overlap with those
of a sample of comparison objects in the IRAC field.
This suggests that IRAC selection may allow detection of
different types of high-redshift galaxies than are found
by Lyman-break selection.
Spectral energy distribution fitting implies that the IRAC-detected 
LBGs are massive stellar systems with relatively recent star 
formation. The sensitivity of IRAC to old stellar light at these high 
redshifts shows its promise for future work in galaxy formation and
evolution.

\acknowledgments

We thank M. Ashby for a careful reading of the manuscript, and the referee
for helpful comments.
This work is based on observations made with the {\it Spitzer Space Telescope},
which is operated by the Jet Propulsion Laboratory, California Institute of 
Technology under NASA contract 1407. Support for this work was provided by NASA 
through Contract Number 1256790 issued by JPL/Caltech.  Support for the IRAC 
instrument was provided by NASA through Contract Number 960541 issued by JPL.
M.A.P. acknowledges NASA/LTSA grant \# NAG5-10777. 

Facilities: Spitzer(IRAC)

The data from this paper, ads/sa.spitzer\#0007127552, ads/sa.spitzer\#0007127808,\\
ads/sa.spitzer\#0007128064, ads/sa.spitzer\#0007128320, ads/sa.spitzer\#0007128576,\\
ads/sa.spitzer\#0007475968, ads/sa.spitzer\#0007476224, ads/sa.spitzer\#0007476480\\
is available through the electornic edition.

\begin{deluxetable}{ccccc}
\tablecaption{IRAC Detections of Lyman break galaxies\label{detnum}}
\tablewidth{0pt}
\tablehead{\colhead{}&\multicolumn{2}{c}{Photometric candidates}
&\multicolumn{2}{c}{Confirmed $z\sim3$}\\
\colhead{IRAC band} & \colhead{N($U_nG\cal R$)\tablenotemark{a}} & \colhead{N(IRAC)\tablenotemark{b}}
& \colhead{N(LRIS)\tablenotemark{a}} & \colhead{N(IRAC)\tablenotemark{b}}}
\startdata
3.6 $\mu$m&114 & 104 & 14 & 14\\
4.5 $\mu$m&180 & 150 & 10 & 10\\
5.8 $\mu$m&112 & 68  & 16 & 14\\
8.0 $\mu$m&170 & 77  & 10 & 7
\enddata
\tablenotetext{a}{Number in field.}
\tablenotetext{b}{Number detected with IRAC.}
\end{deluxetable}

\end{document}